\documentclass[twocolumn,british,pra,aps,showpacs]{revtex4-1}
\usepackage[letterpaper]{geometry}
\usepackage{url}
\usepackage{graphicx}
\usepackage{amsmath}
\usepackage{amssymb}
\usepackage{bbold,soul}
\usepackage{esint}
\usepackage{epstopdf}
\usepackage{subfigure}
\usepackage{color}
\usepackage{soul}
\usepackage{babel}
\newcommand{\bra}[1]{\langle #1|}
\newcommand{\ket}[1]{|#1\rangle}


\definecolor{nblue}{rgb}{0.3,0.3,1.0}
\definecolor{ngreen}{rgb}{0.2,0.7,0.2}
\definecolor{nred}{rgb}{0.9,0.1,0}
\definecolor{norange}{rgb}{0.8,0.5,0}

\begin{document}

\title{On quantum interferometric measurements of temperature}

\author{Marcin Jarzyna}\email{marcin.jarzyna@fuw.edu.pl}
\author{Marcin Zwierz}\email{marcin.zwierz@fuw.edu.pl}
\affiliation{Faculty of Physics, University of Warsaw, ulica Pasteura 5, PL-02-093 Warszawa, Poland}
\date{\today}

\begin{abstract}\noindent
We provide a detailed description of the quantum interferometric thermometer, which is a device that estimates the temperature of a sample from the measurements of the optical phase. For the first time, we rigorously analyze the operation of such a device by studying the interaction of the optical probe system prepared in a single-mode Gaussian state with a heated sample modeled as a dissipative thermal reservoir. We find that this approach to thermometry is capable of measuring the temperature of a sample in the nanokelvin regime. Furthermore, we compare the fundamental precision of quantum interferometric thermometers with the theoretical precision offered by the classical idealized pyrometers, which infer the temperature from a measurement of the total thermal radiation emitted by the sample. We find that the interferometric thermometer provides a superior performance in temperature sensing even when compared with this idealized pyrometer. We predict that interferometric thermometers will prove useful for ultraprecise temperature sensing and stabilization of quantum optical experiments based on the nonlinear crystals and atomic vapors.
\end{abstract}
\maketitle

\section{Introduction}\label{sec:intro}\noindent
Temperature is one of the fundamental and arguably one of the most frequently measured physical quantities. Apart from the central role played by the concept of temperature in the fields of thermodynamics and statistical physics, precise temperature measurements are important for all branches of modern science and technology. Indeed, precise knowledge of the temperature of a sample proved indispensable for many advancements in physics, biology, chemistry, and atmospheric sciences, as well as in material science and the microelectronic industry. In this paper, we study the classical and quantum limitations that constrain the precision with which we can measure temperature. These limitations are imposed by the combination of the laws of statistics, statistical physics, and quantum mechanics.

The theoretical limit on the precision of classical thermometers is known as the standard quantum limit or the shot-noise limit, $\Delta T \sim c_{SQL}/\sqrt{\bar{N}}$ \cite{Giovannetti2011, Demkowicz2015}, where $c_{SQL}$ is a constant depending on the properties of a classical thermometer and the sample, and $\bar{N}$ is the number of resources, which for classical thermometers typically reduces to the (mean) number of \textit{uncorrelated} particles that makes up the thermometer or the time it takes to make a measurement \cite{Stace2010}. The standard quantum limit predicts that in order to measure the temperature of a sample with high precision we need $\bar{N}$ to be large, which translates to enlarging our classical thermometer. A typical measurement consists of bringing the thermometer into physical contact with the sample and then letting the two systems thermalize. However, if we wish to use a large and thus a highly precise thermometer, then this approach is not optimal, as it will, in most cases, significantly disturb the temperature of the sample we are probing. Fortunately, there is a way to avoid this problem. Namely, we could measure the temperature without putting our thermometer and the sample in direct physical contact and thus minimize the sample's temperature disturbance that can arise from the heat exchange. In such a case, the sample of interest is observed remotely. There are many different techniques of the so-called noninvasive or noncontact thermometry that are used across many branches of science and industry. Most noninvasive techniques infer temperature from the electromagnetic spectrum. One of the most common optical techniques that is used for temperature measurements is the measurement of the thermal infrared radiation naturally emitted by all heated samples \cite{Childs2000, Blackburn2004}. This is exactly the principle used in the commercially available pyrometers.

On the other hand, recent years have witnessed a growing interest in applying various ideas that take advantage of quantum mechanical features of nature to the problem of temperature measurements \cite{Stace2010, Brunelli2011, Brunelli2012, Higgins2013, Marzolino2013, Martin-Martinez2013, Jevtic2015, Correa2014, Weng2014}. Specifically, in Ref.~\cite{Stace2010}, it was shown that it is possible to map the problem of measuring the temperature onto the problem of estimating an unknown phase. Moreover, it was further shown that by using the techniques of phase estimation theory, the classical standard-quantum-limited precision in temperature estimation may be improved to the so-called Heisenberg limit, $\Delta T \sim c_{HL}/\bar{N}$ \cite{Holland1993, Ou1997, Giovannetti2006, Zwierz2010, Braun2012, Hall2012, Hall2012pi}, where $c_{HL}$ is a constant depending on the properties of a quantum thermometer and the sample, and $\bar{N}$ is the number of resources, which for quantum thermometers usually denotes the (mean) number of \textit{correlated} particles that makes up the thermometer \cite{Stace2010}. Unfortunately, this improvement occurs only for measurements in the absence of decoherence; typically, the presence of noise reduces the Heisenberg-limited precision back to the classical shot-noise-limited scaling, $\Delta T \sim c/\sqrt{\bar{N}}$, with a possible advantage constrained to the scaling constant $c$, where $c < c_{SQL}$ \cite{Escher2011,Demkowicz2012}. Nevertheless, the prospect of using interferometric tools of quantum-enhanced phase estimation for temperature measurements is rather intriguing and, we believe, has not been fully explored.

The main advantages of interferometric thermometry are a very fast response time of interferometric devices, which means that rapid variations in temperature can be measured, the ability to measure a wide range of temperatures, and the high spatial resolution allowing for temperature measurements of micrometer- and even nanometer-size spots \cite{Childs2000, Blackburn2004}. When combined in a single measurement setup, these advantages open up a possibility of preparing temperature maps with very high spatial and temporal resolutions that could find a wide range of important applications in the microelectronic industry, and in the material and life sciences. Surprisingly, in spite of those most immediate benefits, interferometric thermometry met with limited attention in the past. Interferometric thermometry was used mostly in the microelectronic industry for temperature measurements of the semiconductor electronic devices; in industry, this approach to thermometry is used not only for local temperature measurements but also as a kind of scanning device for measuring the thermal expansion coefficient \cite{Breuer2005, Childs2000}. In academia, apart from the theoretical proposal discussed above, the idea of interferometric thermometry realized with atomic quantum dots was most recently employed to study temperature measurements of the Bose-Einstein condensates \cite{Sabin2014}.

To the best of our knowledge, this is the first work that rigorously studies the interferometric approach to thermometry. We determine the fundamental precision of optical interferometric thermometers. To this end, we use the single-mode Gaussian states of light to probe the temperature variations in a general medium modeled as a dissipative thermal reservoir. We then compare our precision limits against the theoretical limit on the precision offered by commonly used classical noninvasive temperature sensors, the so-called pyrometers. This comparison is provided for one of the most important classes of materials used in many quantum technologies, namely, the nonlinear crystals. A precise sensing of tiny temperature variations followed by an active temperature stabilization is required in many experimental setups based on the nonlinear crystals such as the PPKTP crystal. The nonlinear crystals are used as a source of indistinguishable photons generated via the spontaneous parametric down-conversion process and the quality of those photons depends strongly on the temperature of the crystal \cite{Steinlechner2013, Jachura2014}. Hence, new practical methods for ultraprecise temperature sensing and stabilization, which could be easily incorporated into the existing experimental setups, are of paramount importance for the development of various optical quantum technologies that rely heavily on the bright sources of truly indistinguishable photons. The interferometric approach to thermometry that we study appears ideal for such applications.

The paper is organized as follows. In Sec.~\ref{sec:estimation}, we introduce the basic concepts of the classical and quantum estimation theory. In Sec.~\ref{sec:pyrometer}, we derive the fundamental precision limits for the idealized classical noninvasive pyrometers that we use as a benchmark to judge the usefulness of the interferometric thermometry. Section \ref{sec:interthermo} introduces in detail the idea of interferometric thermometry. In Sec.~\ref{sec:phase}, we find the precision of temperature estimation obtained via the interferometric phase measurement in the presence of a dissipative thermal reservoir focusing on temperature measurements in nonlinear crystals. In Sec.~\ref{sec:conclusions}, we conclude with final remarks.

\section{Classical and quantum estimation theory}\label{sec:estimation}
\begin{figure}[t!]
\centering
\includegraphics[width=7.8cm]{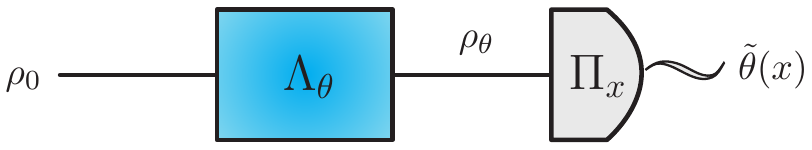}
\caption{(Color online) A general parameter estimation scheme in which a probe system prepared in an initial quantum state $\rho_{0}$ evolves under the quantum channel $\Lambda_\theta$ to the output quantum state $\rho_\theta$. Note that $\Lambda_\theta$ takes into account all decoherence processes that may be present in the setup. Following the evolution stage, the probe system is subjected to a general quantum measurement described by a POVM $\{\Pi_{x}\}$, which produces the measurement results $x$. These measurement results are then used to make an estimate of the parameter $\theta$ via the estimator function $\tilde{\theta}(x)$.}
\label{fig:generalscheme}
\end{figure}

\noindent
The task of temperature measurement can be translated applying the language of estimation theory to the problem of parameter estimation \cite{Giovannetti2011, Demkowicz2015}. The generic scheme for such a problem is depicted in Fig.~\ref{fig:generalscheme}. In order to estimate the value of parameter $\theta$, we send a probe system prepared in an initial quantum state $\rho_0$ through a sample, which depends on $\theta$. The interaction between our probe system and the sample can then be modeled as an evolution under a quantum channel $\Lambda_\theta$. Hence, following the interaction, the probe system is left in the output quantum state $\rho_\theta = \Lambda_\theta[\rho_0]$. Next, we subject the probe system to a general quantum measurement, described by a positive operator-valued measure (POVM) $\{\Pi_x\}$, which produces the measurement results $x$. Finally, we use a special function $\tilde{\theta}(x)$ called an estimator to calculate the estimated value of $\theta$. This scheme describes not only the quantum estimation tasks, but also the classical ones \cite{Helstrom1976, Holevo1982}.

The procedure described above naturally never returns the true value of $\theta$. We quantify the discrepancy between the estimated value and the true value of $\theta$ with the root-mean-square error $\Delta\theta = [\langle(\tilde{\theta}(x)-\theta)^2\rangle]^{1/2}$, where the average is taken with respect to the probability distribution $p(x|\theta)$ of the measurement outcomes $x$. In the most common case when we have a specific POVM measurement $\{\Pi_x\}$, corresponding to an observable $A$, and we wish to estimate the value of $\theta$ from the mean value of $A$, the precision is given by the standard error propagation formula \cite{Helstrom1976, Holevo1982},
\begin{equation}\label{eq:propagation}
\Delta \theta = \frac{\Delta A}{|d\langle A\rangle / d \theta|},
\end{equation}
where $\Delta A$ is the standard deviation of the observable $A$ calculated for the output state $\rho_\theta$. We remind the reader that $A$ may as well be a classical observable.

The above equation is valid for any observable, but it explicitly assumes an estimation from the mean value of $A$. It may happen that this is not the optimal approach and we could improve the precision by choosing another estimator. Moreover, it may also happen that the measurement we choose is not optimal either. In order to avoid such problems, we need to optimize Eq.~(\ref{eq:propagation}) over all possible measurements and estimators, which is not an easy task. Fortunately, the solution to this optimization problem is provided by the classical and quantum Cram\'{e}r-Rao inequalities \cite{Helstrom1976, Holevo1982, Braunstein1994}, which state that for any \textit{unbiased} estimator,
\begin{equation}\label{eq:CRI}
\Delta \theta \geq \frac{1}{\sqrt{k F_{\theta}}} \geq \frac{1}{\sqrt{k Q_{\theta}}},
\end{equation}
where $k$ is the number of independent experimental repetitions and $F_{\theta}$ is the classical Fisher information defined as \begin{equation}\label{eq:Fclas}
F_{\theta} = \sum_{x} \frac{1}{p(x)} \left[ \frac{\partial p(x)}{\partial \theta} \right]^2.
\end{equation}
The classical Fisher information gives a lower bound on the precision optimized over all unbiased estimators for a specific POVM measurement $\{\Pi_x\}$. Further optimization over all conceivable measurements gives a lower bound on the precision expressed via the quantum Fisher information (QFI) $Q_\theta$, which can be computed from $Q_\theta={\rm Tr}[\rho_{\theta} L_\theta^2]$, where $L_\theta$ is a Hermitian operator called the symmetric logarithmic derivative (SLD) that is implicitly defined via
\begin{equation}
\frac{d\rho_{\theta}}{d\theta} = \rho_{\theta} \circ L_\theta,
\end{equation}
where we used the notation of the symmetric product for operators, $A \circ B = \frac{1}{2}(AB + BA)$. In a situation where the parameter is encoded by a unitary transformation, i.e., $\rho_\theta=\Lambda_\theta[\rho_0]=U_\theta\Lambda[\rho_0]U_\theta^\dagger$, where $U_\theta$ is unitary and $\Lambda$ is a $\theta$-independent quantum channel describing possible decoherence processes, then $L_\theta=U_\theta L_0U_\theta^\dagger$, where $L_0$ is the SLD for the state $\Lambda[\rho_0]$. In such a case, the QFI does not depend on $\theta$, that is, $Q_\theta = Q$.

The quantum Cram\'{e}r-Rao bound (QCRB) and its classical counterpart are known to be saturable in the limit of a large number of repetitions, $k\to\infty$. The optimal measurement consists of a set of projectors on the eigenbasis of the SLD and outputs the measurement results, which are then processed with the maximum-likelihood estimator \cite{Giovannetti2011, Demkowicz2015}. Such projective measurements are typically very hard to implement reliably in the laboratory; however, very often it is possible to saturate the quantum Cram\'{e}r-Rao bound with a more natural set of measurements, for example, with the optical homodyne or the parity detection \cite{Ono2010, Seshadreesan2011, Demkowicz2015}. Since the QCRB is already optimized over all possible POVMs and unbiased estimators, we are usually only left with the optimization over all possible input quantum states $\rho_{0}$.

\section{Noninvasive classical thermometers}\label{sec:pyrometer}\noindent
As we have already mentioned, one of the most widely used optical noninvasive temperature sensors relies on a measurement of thermal infrared radiation naturally emitted by all heated samples \cite{Childs2000, Blackburn2004}. This measurement principle is used in the commercially available pyrometers. Before we explain the operation of these devices in detail, let us explain what we understand by the term noninvasive. Pyrometers measure the temperature of a sample by probing the thermal radiation that the sample emits, but not the sample itself. This thermal radiation and the sample are in thermal contact, resulting in a thermal equilibrium state. In the following, we assume that the presence of the pyrometer (and a short probing time) has a negligible influence on this equilibrium state, and hence, in this sense, we can consider pyrometers as noninvasive.

What is the fundamental precision of this type of temperature sensors? Pyrometers estimate temperature by measuring the flux of thermal radiation emitted by a sample modeled as a blackbody in thermal equilibrium, and using the well-known Stefan-Boltzmann law, infer the temperature \cite{Childs2000, Blackburn2004}. Typically, those devices take into account the nonunit emissivity $\varepsilon$ of most physical samples; however, for the purpose of this derivation, we assume that we are dealing with a perfect blackbody. The flux of thermal radiation $\Phi$, that is, the total energy radiated per unit surface area of a blackbody across all wavelengths per unit time, is quantified by the Stefan-Boltzmann formula,
\begin{equation}
\Phi = \sigma T^4 \quad {\rm with} \quad \sigma=\frac{\pi^2k_B^4}{60\hbar^3c^2},
\end{equation}
where $\sigma$ is the Stefan-Boltzmann constant and $T$ is the temperature of the sample. In order to calculate the precision of temperature estimation based on the Stefan-Boltzmann law, we use the error propagation formula given in Eq.~(\ref{eq:propagation}) with $\langle A \rangle = \Phi$. Using the definition of the variance in the thermal photon number $(\Delta N)^2 = N(N + 1)$, with $N$ being the mean number of thermal photons distributed according to the Bose-Einstein statistics \cite{Barnett2003},
\begin{equation}\label{eq:meanthermalnumber}
N = \left[\exp\left(\frac{\hbar\omega}{k_{B} T}\right) - 1\right]^{-1},
\end{equation}
and assuming for simplicity that pyrometers are sensitive to all frequencies of electromagnetic radiation \footnote{In fact, the commercially available pyrometers operate in a surprisingly narrow band of electromagnetic spectrum, extending from approximately 0.7 to 20 $\mu$m, because these devices are not sensitive enough beyond this band. However, for most common applications, this range is perfectly sufficient as most of the thermal radiation is emitted in the range of 0.1--100 $\mu$m \cite{Childs2000}.} and excluding all sources of loss in the detection process, we can calculate
\begin{equation}
\Delta \Phi = \sqrt{\int_{0}^{\infty} (\hbar \omega \Delta N)^2 \wp(\omega) d\omega} = \sqrt{4 k_{B} \sigma T^{5}},
\end{equation}
where $\wp(\omega) = \omega^2/(4\pi^2c^2)$ is the density of modes per unit interval in $\omega$ in a unit area. This allows us to find the fundamental precision in temperature estimation for the idealized pyrometer. However, the error that we obtain from the error propagation formula represents the square root of the inverse of the information collected per unit surface area, per unit measurement time. In order to include the total information collected by the pyrometer, we need to divide this error by $\sqrt{S \delta t}$, where $S$ is the surface area of the sample we are probing, which we assume is optimized to be equal to the size of the detector area used in our pyrometer, and $\delta t$ is the response time of the device, that is, the time it takes to make a measurement. This leads to the following theoretical limit on the precision in temperature estimation for the idealized pyrometer:
\begin{equation}\label{eq:pyrometer}
\Delta T = \sqrt{\frac{k_B}{4\sigma S \delta t T}}.
\end{equation}
This result has a standard-quantum-limited scaling with respect to the response time $\delta t$, which here corresponds to the amount of resources $\bar{N}$ we discussed in Sec.~\ref{sec:intro} \footnote{The term standard quantum limit is always used in relation to classically limited measurements and it has always been considered as an oxymoron. However, here the quantum character of this scaling is fully justified because the scaling constant $c_{SQL} = \sqrt{k_B/(4\sigma S T)}$ originates from the Stefan-Boltzmann law, which represents one of the very first results of quantum theory.}; therefore, the longer we probe the sample, the more photons the detector registers. Modern pyrometers operate in the ms regime; hence, we assume $\delta t = 10$ ms. Now further assuming that we measure the temperature of a surface with $S = 1$ cm$^2$, we find that the local precision of such a measurement near $T$ = 298 K is $\Delta T = 452$ nK. Therefore, at the fundamental level, pyrometers are highly precise devices. Naturally, a real commercially available pyrometer estimates temperature with a much lower precision; a typical precision at the room-temperature range is, at best, 0.1 K.

Is the measurement of the total thermal radiation the most optimal measurement we can perform? Surely, this is not the only way we could use thermal light to infer the temperature of a sample. We could, for example, imagine a more sophisticated device that estimates temperature based on the detailed knowledge of the spectral photon-number distribution of thermal radiation. Such a device would measure the photon-number probability distribution for each frequency. This kind of measurement would provide more information about the character of the thermal light and, therefore, should allow for improved temperature estimation. In order to calculate this enhanced precision, we need to find the classical Fisher information for a photon-number-resolving measurement. This is a measurement that allows us to estimate the photon-number probability distribution $p(n)$, which enters into the definition of the thermal state,
\begin{equation}\label{eq:StateThermal}
\rho_{N}=\sum_{n=0}^{\infty} p(n) \ket{n}\bra{n} \quad {\rm with} \quad p(n) = \frac{N^n}{(1+N)^{n+1}},
\end{equation}
where $N$ is the mean number of photons present in the thermal state given in Eq.~(\ref{eq:meanthermalnumber}). The classical Fisher information for $p(n)$ for a single frequency $\omega$ is
\begin{equation}\label{eq:singleFisher}
F_T(\omega) = \left(\frac{\hbar \omega}{k_B T^2}\right)^2 N(N + 1).
\end{equation}
Surprisingly, when we calculate the total Fisher information per unit area for all frequencies,
\begin{equation}\label{eq:qfiThermal}
F_T = \int_{0}^{\infty} F_{T}(\omega) \wp(\omega) d\omega,
\end{equation}
where $\wp(\omega) = \omega^2/(4\pi^2c^2)$ is the density of modes per unit interval in $\omega$ in a unit area, we obtain via the classical Cram\'{e}r-Rao inequality the very same expression as in Eq.~(\ref{eq:pyrometer}), that is,
\begin{equation}\label{eq:pyrometerFcl}
\Delta T \geq \frac{1}{\sqrt{\delta t S F_{T}}} =  \sqrt{\frac{k_B}{4\sigma S\delta t T}}.
\end{equation}
This result means that inferring temperature from the total flux of thermal radiation is already optimal. Any additional knowledge of the photon-number distribution does not contribute to the information about temperature. Furthermore, as can be easily checked, the calculation of the QFI for $\rho_N$ returns the same result because the thermal state is diagonal in the Fock basis and so is the corresponding SLD \cite{Stace2010, Nair2015}. Therefore, it is always optimal to measure the mean number of photons at the output for each frequency, as we have effectively done in the derivation of Eq.~(\ref{eq:pyrometerFcl}), or simply to measure the total flux of thermal radiation.

\section{Quantum interferometric thermometers}\label{sec:interthermo}
\begin{figure}[t!]
\centering
\includegraphics[width=7.8cm]{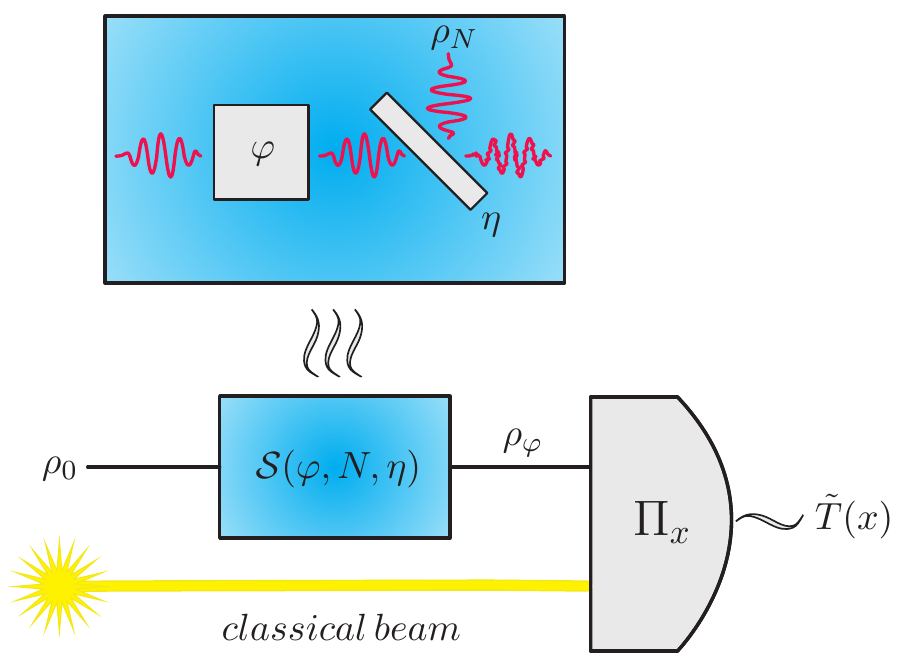}
\caption{(Color online) Basic scheme for quantum interferometric thermometry. A single-mode Gaussian state of light $\rho_0$ is sent through a sample with temperature $T$ and transmissivity $\eta$, resulting in a mixed output state $\rho_\varphi$, which is then measured using a general POVM $\{\Pi_x\}$ measurement. Assuming that the phase shift $\varphi$ is temperature dependent, the measurement outcomes are then used to find an estimated value of $T$ via the estimator function $\tilde{T}(x)$. The bright yellow beam depicts a classical reference, which allows us to define the phase shift in a meaningful way \cite{Jarzyna2012}. The propagation of light through a heated sample with transmissivity $\eta$ can be decomposed into two distinct processes: first the single-mode Gaussian beam undergoes the phase shift $\varphi$ relative to the reference beam and then it undergoes a photon loss process, which is modeled with the help of a beam splitter with transmissivity $\eta$, where the second port is filled with light in the thermal state $\rho_N$.}
\label{fig:scheme}
\end{figure}

\noindent
Having explained how the noninvasive classical thermometers work, we now describe in detail the operation of quantum interferometric thermometers. It may be somewhat confusing to consider the interferometric thermometry as noninvasive or noncontacting since, as will be apparent soon, the photons in the probe beam are clearly interacting with the sample and, therefore, the thermometer is in direct physical contact with the sample, implying an exchange of energy \footnote{Because of this, our realistic interferometric thermometer does not exactly belong to the class of nonthermalizing thermometers that was considered in Ref.~\cite{Stace2010}. On the other hand, it does not belong to the class of thermalizing thermometers either, as it does not attain the thermal equilibrium with the sample}. However, as long as this interaction and the ensuing exchange of energy do not cause a significant change in the sample's temperature, we can consider this method as truly noninvasive \cite{Blackburn2004}. The scope of this work is to investigate a possible advantage offered by such interferometric devices that typically rely on the use of nonclassical states of light. The basic scheme for a quantum interferometric thermometer is depicted in Fig.~\ref{fig:scheme}. A single-mode Gaussian state of light prepared in an initial state $\rho_{0}$ propagates through a sample with temperature $T$ and transmissivity $\eta$, resulting in a mixed output state $\rho_\varphi$, where $\varphi$ is the temperature-dependent phase shift that we wish to estimate. Following the propagation stage, a general quantum measurement described by a POVM $\{\Pi_x\}$ is performed on the output state $\rho_\varphi$, returning a measurement outcome $x$, which is then used to estimate the value of $T$ via the estimator $\tilde{T}(x)$. For the sake of simplicity, we consider here only single-mode Gaussian states of light; however, we note that a proper interferometric setup requires a reference, which in Fig.~\ref{fig:scheme} is depicted as a bright classical beam.

\subsection{Input state}\noindent
We choose to work with the single-mode Gaussian states, which include the single-mode coherent and squeezed states, because this particular class of states is readily available with the current technology \cite{Adesso2006, Weedbrook2012}. The most general single-mode Gaussian state of light can be parametrized as
\begin{equation}
\rho_{0} = D(\alpha_{0}) S(r_{0}) \rho_{N_{0}} S^{\dagger}(r_{0}) D^{\dagger}(\alpha_{0}),
\end{equation}
where $\rho_{N_{0}}$ is a single-mode thermal state with mean photon number $N_{0}$, which is formally defined in Eq.~(\ref{eq:StateThermal}), $S(r_{0}) = \exp\left[\frac{1}{2}(r_{0} a^{\dagger 2} - r^{*}_{0} a^{2})\right]$ is the squeezing operator, and $D(\alpha_{0}) = \exp\left[\alpha_{0} a^{\dagger} - \alpha^{*}_{0} a \right]$ is the displacement operator, with $a$ and $a^{\dagger}$ being, respectively, the annihilation and creation operators of a bosonic mode. Alternatively, we may express the above operators in terms of the canonical position and momentum operators $x$ and $p$, which can be arranged into a row vector ${\bf d} = (x, p)$. For example, we can express the displacement operator as $D({\bf d}_{0}) = \exp\left[i(\bar{p}_{0} x - \bar{x}_{0} p)\right]$ with ${\bf d}_{0} = (\bar{x}_{0}, \bar{p}_{0})$.

It is a well-known fact that a Gaussian state is fully characterized by only its first and second canonical moments \cite{Weedbrook2012},
\begin{equation}
W({\bf d}) = \frac{\exp\left[-\frac{1}{2}({\bf d} - {\bf d}_{0})^{\intercal}\Sigma_{0}^{-1}({\bf d} - {\bf d}_{0})\right]}{2\pi \sqrt{{\rm det} \Sigma_{0}}},
\end{equation}
where $W({\bf d})$ is the Wigner quasiprobability distribution of a single-mode Gaussian state. The first moments are defined via ${\bf d}_{0}$. The second moments can be arranged in the so-called covariance matrix, which for the single-mode state $\rho_{0}$ is given by
\begin{equation}
\Sigma_{0} = \left(N_{0} + \frac{1}{2}\right) \left(\begin{array}{cc}
e^{2 r_{0}} & 0\\
0 & e^{-2 r_{0}}
\end{array}\right).
\end{equation}
The mean number of photons $\bar{N}$ in such an input Gaussian state is given by
\begin{equation}
\bar{N} = \frac{1}{2}\left[\left(N_{0} + \frac{1}{2}\right) 2 \cosh 2r_{0} + \bar{x}_{0}^{2} + \bar{p}_{0}^{2} - 1\right].
\end{equation}
In the next section, we describe how the propagation through a sample with temperature $T$ and transmissivity $\eta$ affects the above single-mode Gaussian state.

\subsection{Evolution and the output state}\noindent
The evolution of an arbitrary state $\rho$ in the presence of a dissipative thermal reservoir is described by the following master equation:
\begin{equation}\label{eq:master}
\frac{d \rho}{d t} = \mathcal{G}(\omega, N, \Gamma)\rho
\end{equation}
with the superoperator $\mathcal{G}(\omega, N, \Gamma)$ defined as
\begin{equation}
\mathcal{G}(\omega, N, \Gamma) = - i \omega H + \frac{\Gamma}{2}(N L[a^{\dagger}]+(N + 1)L[a]),
\end{equation}
where $H\rho=[a^{\dagger}a,\,\rho]$ and $L[o]\rho=2 o\rho o^{\dagger}-o^{\dagger}o\rho-\rho o^{\dagger}o$. The first term in the superoperator $\mathcal{G}$ describes a free unitary evolution of a single bosonic mode $a$ with frequency $\omega$. [Naturally, the frequency appearing in the definition of $N$ in Eq.~(\ref{eq:meanthermalnumber}) is equal to the frequency $\omega$ of the incident Gaussian light.] The second term accounts for a coupling of the bosonic mode to a thermal reservoir (with mean photon number $N$) with strength $\Gamma$ \cite{Barnett2003}.

According to the above master equation, after a time $t$, an initial state $\rho_{0}$ evolves to $\rho_\varphi = \exp[\mathcal{G}(\omega, N, \Gamma) t]\rho_{0} = \mathcal{S}(\varphi, N, \gamma) \rho_{0}$, where $\mathcal{S}(\varphi, N, \gamma) = \exp[\mathcal{G}(\varphi, N, \gamma)]$, and here $\gamma = \Gamma t$ is the effective coupling to a thermal reservoir (which is related to the photon loss coefficient $\eta$ via the relation $\eta = e^{-\gamma}$) and $\varphi = \omega t$ is the temperature-dependent optical phase that we are interested in estimating.

Now, when the initial state $\rho_{0}$ corresponds to our initial single-mode Gaussian state, then following the above evolution, the output state $\rho_\varphi$ is still a single-mode Gaussian state but with the changed first and second moments. By transforming the master equation in Eq.~(\ref{eq:master}) into a partial differential equation, the so-called Fokker-Planck-type equation, for the Wigner quasiprobability distribution $W(\alpha, \alpha^{*}, t)$ \cite{Barnett2003},
\begin{eqnarray}
\frac{\partial W}{\partial t} &=& \left(\frac{\Gamma}{2} + i \omega \right) \frac{\partial}{\partial \alpha}[\alpha W] + \left(\frac{\Gamma}{2} - i \omega \right) \frac{\partial}{\partial \alpha^{*}}[\alpha^{*} W] \nonumber \\
&+& \Gamma \left(N + \frac{1}{2}\right)\frac{\partial^{2} W}{\partial \alpha \partial \alpha^{*}}
\end{eqnarray}
and then solving this equation for the complex variables $\alpha$ and $\alpha^{*}$ \cite{Barnett2003}, we can show that the first moments of the output state $\rho_\varphi$ are given by
\begin{eqnarray}\label{eq:meanx}
\bar{x} &=& \sqrt{\eta}(\cos\varphi \, \bar{x}_{0} + \sin\varphi \, \bar{p}_{0}), \\
\bar{p} &=& \sqrt{\eta} (-\sin\varphi \, \bar{x}_{0} + \cos\varphi \, \bar{p}_{0}). \label{eq:meanp}
\end{eqnarray}
The second moments of the output state are given by the following covariance matrix:
\begin{equation}\label{eq:Sigma}
\Sigma =  \Sigma_{\eta}(\Sigma_{\varphi} - \Sigma_{N})\Sigma_{\eta} + \Sigma_{N},
\end{equation}
where $\Sigma_{\eta}=\sqrt{\eta}\mathbb{1}$, $\Sigma_{N}=(N+\frac{1}{2})\mathbb{1}$, and $\Sigma_{\varphi} = R(\varphi) \Sigma_{0} R^{\intercal}(\varphi)$ is the covariance matrix $\Sigma_{0}$ of the input state $\rho_{0}$ rotated by an angle $\varphi$ \cite{Weedbrook2012}. Because in our setup the optical phase $\varphi$ is encoded by a unitary transformation (as $H$ commutes with $L[a]$ and $L[a^{\dagger}]$), the QFI for $\varphi$ will not depend on the actual value of $\varphi$. Therefore, in the remaining sections, we always neglect the rotation about $\varphi$ by setting $\varphi=0$.

The above analysis shows that the propagation of light through the sample may be modeled as a Gaussian channel, which may be effectively decomposed into two distinct processes: first, the single-mode Gaussian beam acquires the phase shift $\varphi$ relative to the reference beam and then it undergoes a photon loss process, which is modeled with the help of a beam splitter with transmissivity $\eta$, where the second port is filled with thermal light $\rho_{N}$ with temperature $T$. Physically, (the temperature-dependent part of) the acquired phase shift is caused by the thermal expansion of a sample and small temperature-dependent changes in the refractive index $n$. Assuming that we probe a sample with a length $L$, the refractive index $n$, the thermo-optic coefficient $n'=dn/dT$, and the thermal expansion coefficient $\alpha_T$, all known exactly for a specific temperature $T$ \footnote{We note that the systematic error of the thermometer associated with the uncertainties of those physical parameters would need to be estimated when calibrating the device for a specific temperature range.} with light with frequency $\omega$, the acquired phase shift is described by the simple relation $\varphi = n(\delta T)\omega L(\delta T)/c$, where $n(\delta T) = n + n' \delta T$ and $L(\delta T) = L(1 + \alpha_{T} \delta T)$. The temperature-dependent part of the acquired phase shift given to the first order in $\delta T$ can be written as
\begin{equation}
\varphi = \frac{\omega L}{c}(n\alpha_T+n') \delta T = \alpha T,
\end{equation}
where for the sake of simplicity we replaced $\delta T$ with $T$. However, we emphasize that in this work, we always estimate or probe tiny deviations of temperature $\delta T$ from a known value.

In the following section, we determine the best possible precision with which we can estimate the temperature of a nonlinear PPKTP crystal with the interferometric phase measurement.

\section{Temperature estimation with application to nonlinear crystals}\label{sec:phase}\noindent
We now wish to calculate the QFI for temperature. To this end, we need to first calculate the QFI for the optical phase $\varphi$. In general, any calculation involving a single-mode Gaussian beam interacting with a dissipative thermal reservoir is very complicated, to the point where, in many cases, only numerical results can be obtained. The calculation of the QFI for the optical phase shift acquired in such a setup is no exception. Fortunately, in Ref.~\cite{Monras2010}, the authors developed a very powerful technique for the derivation of the SLD and the QFI for arbitrary Gaussian probe states propagating through general dissipative Gaussian reservoirs. Here, we adopt this technique to find the SLD and the QFI for the optical phase. [We should mention the two related results presented in Refs.~\cite{Pinel2013} and \cite{Monras2013} that can also be used to find the SLD and the QFI for the optical phase but these results were obtained using different methods.] The details of the derivation of the SLD for the optical phase given in terms of the first and second moments of the output state $\rho_{\varphi}$ are presented in the Appendix. Here we only present the resulting QFI,
\begin{equation}\label{eq:QFIphase}
Q_{\varphi} = \frac{4 (\Sigma_{22} - \Sigma_{11})^2}{1 + 4\Sigma_{11} \Sigma_{22}} + \frac{\bar{x}^2}{\Sigma_{22}} + \frac{\bar{p}^2}{\Sigma_{11}},
\end{equation}
where $\Sigma_{11}$ and $\Sigma_{22}$ are the diagonal elements of the covariance matrix $\Sigma$ of the output state $\rho_{\varphi}$, and $\bar{x}$ and $\bar{p}$ are the mean displacements in the respective canonical position and momentum quadratures, all of which are calculated with $\varphi = 0$. Based on the above formula for $Q_{\varphi}$ and using a simple reparametrization, we can easily find the following QFI for temperature:
\begin{equation}\label{eq:FisherT}
Q_T = \left(\frac{d \varphi}{d T}\right)^2 Q_{\varphi}.
\end{equation}

We now focus on finding the maximum value of $Q_{\varphi}$ (and, by implication, the maximum value of $Q_{T}$) and the optimal state that asymptotically attains this value. To this end, we need to optimize Eq.~(\ref{eq:QFIphase}) over the input state parameters, that is, optimize over the mean displacements $\bar{x}_{0}$ and $\bar{p}_{0}$, the mean number of thermal photons $N_{0}$, and the amount of squeezing $r_{0}$. In the limit of a large average number of input photons $\bar{N} \gg 1$, the numerical optimization of Eq.~(\ref{eq:QFIphase}) predicts that the optimal input state is a squeezed-vacuum state, which implies $\bar{x}_{0} = \bar{p}_{0} = 0$ and $N_{0} = 0$. Hence, for the squeezed-vacuum input state, the asymptotic error of temperature estimation, which holds in the limit of large $\bar{N}$, as given by the Cram\'{e}r-Rao inequality, reads
\begin{equation}\label{eq:QFIboundSq}
\Delta T \geq \frac{1}{\alpha} \sqrt{\frac{(1-\eta)(1+2N)}{4\eta\bar{N}}} + \mathcal{O}\left[\frac{1}{\bar{N}}\right].
\end{equation}
This bound scales as $c_{SQL}/\sqrt{\bar{N}}$. The readers familiar with the problem of phase estimation in the presence of dissipative reservoirs will certainly notice that the above bound, neglecting for a moment the prefactor $1/\alpha$, resembles a bound that is typically obtained for phase estimation in the presence of photon loss \cite{Escher2011, Demkowicz2012}. However, our bound has an additional coefficient $\sqrt{1+2N}$ because we consider here a dissipative reservoir prepared in a thermal state $\rho_{N}$, whereas in the lossy phase estimation, it is more common to assume a reservoir prepared in the vacuum state \cite{Demkowicz2009}. In fact, it would be very interesting to determine what kind of error scaling can be obtained for phase estimation when the reservoir is prepared in a pure squeezed state \cite{Jarzyna2015}. We further note that for visible light and moderate temperatures, the mean number of thermal photons $N$ is very small. However, we include $N$ in our analysis because we wish to present a general and realistic model of a quantum interferometric thermometer, which can be applied to all relevant ranges of electromagnetic spectrum such as, the microwave range, and all relevant physical systems, including systems for which the mean number of thermal \textit{excitations} (not necessarily photons) is sizable such as the mechanical systems \cite{Aspelmeyer2014}.

It is also instructive to determine how well we can estimate temperature if we use coherent states (implying $N_{0} = 0$ and $r_{0} = 0$) as the input states to our setup instead of the optimal squeezed-vacuum states. In this case, the Cram\'{e}r-Rao inequality predicts that the asymptotic error, which holds in the limit of large $\bar{N}$, is lower bounded by
\begin{equation}\label{eq:QFIboundCoh}
\Delta T \geq \frac{1}{\alpha} \sqrt{\frac{1+2(1-\eta)N}{4\eta\bar{N}}} + \mathcal{O}\left[\frac{1}{\bar{N}}\right],
\end{equation}
which for a zero-temperature reservoir, that is, $N = 0$, and neglecting the prefactor $1/\alpha$ is equivalent to the well-known bound in lossy phase estimation \cite{Demkowicz2015}.

\subsection{Noninvasiveness of quantum interferometric thermometers}\noindent
From the above bounds, we could draw a conclusion that it is best to send as much light into the sample as possible because with increasing $\bar{N}$, the error in temperature estimation becomes smaller. However, if we were to use very strong squeezed-vacuum or coherent states, then because of the absorption of light this would at some point disturb the sample's temperature. We take this effect into account by adding the magnitude of this disturbance to our precision bound. If the sample absorbs, on average, $N_{\rm abs} = (1 - \eta)\bar{N}$ photons, then its temperature is \textit{at worst} disturbed by $\delta T = \hbar\omega N_{\rm abs}/(M C_{s})$, where $M$ and $C_s$ are the mass and the specific heat of the sample, respectively. Hence, as long as $\delta T \ll \Delta T$, the sample's temperature is not disturbed significantly and our interferometric thermometer is noninvasive.

\begin{figure}[t!]
\centering
\includegraphics[width=7.8cm]{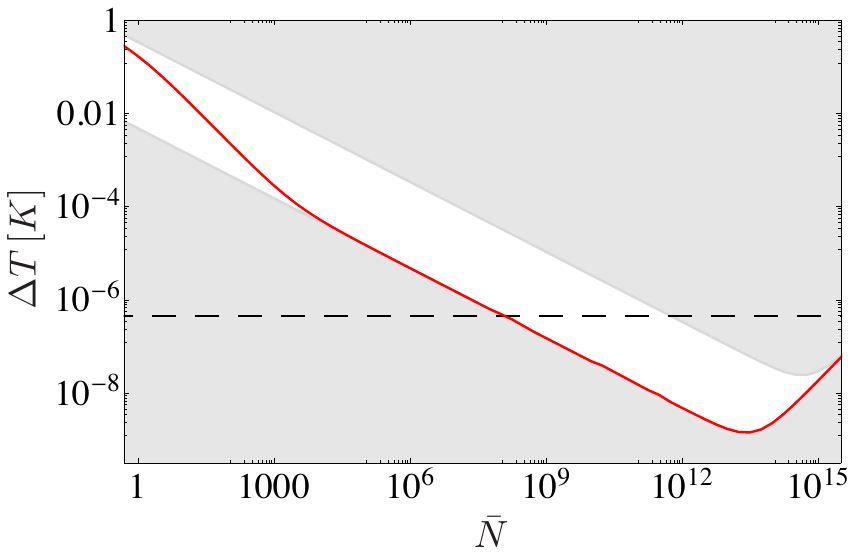}
\caption{(Color online) The precision of the quantum interferometric thermometer near $T=298$ K for a nonlinear PPKTP crystal with a length of $1$ cm plotted as a function of the mean number of photons $\bar{N}$ with $\lambda=1064$ nm ($\omega=1.77 \times 10^{15}$ Hz). In our model for the PPKTP crystal, we set $M = 3$ g, $C_{s} = 688$ $\textrm{J(kgK)}^{-1}$, $n = 1.74$, $\alpha_T = 1.1 \times 10^{-5}$ $\textrm{K}^{-1}$, $n' = 0.6 \times 10^{-5}$ $\textrm{K}^{-1}$, and $\eta = \exp[-L \alpha_{\textrm{abs}}] = 0.9998$, with $\alpha_{\textrm{abs}} = 0.0002$ cm$^{-1}$ \cite{Nikogosyan2005}. The red solid curve represents the \textit{exact} lower bound for the optimal single-mode Gaussian state obtained via the QCRB with $k=1$ and $Q_{T}$ given in Eq.~(\ref{eq:FisherT}), with the tail correcting for a possible heating up of the crystal. The gray areas depict the precision regions lying below the asymptotic bound for the single-mode squeezed-vacuum state given in Eq.~(\ref{eq:Tpopr}) and lying above the asymptotic bound for the single-mode coherent state given in Eq.~(\ref{eq:Tcohpopr}). The black dashed line represents the fundamental precision for the idealized pyrometer given in Eq.~(\ref{eq:pyrometer}) with $S = 1$ $\textrm{cm}^2$ and $\delta t = 10$ ms.}
\label{fig:bounds}
\end{figure}

The effect of disturbance can be included into our asymptotic error bounds by adding $\delta T$ on the right-hand side of Eqs.~(\ref{eq:QFIboundSq}) and (\ref{eq:QFIboundCoh}). We note that by doing so we are combining in a single formula two different types of errors: the statistical error and the systematic error associated with the heating of the sample. However, this allows us to depict both of these errors neatly in the same figure. The resulting formulas for the asymptotic error of temperature estimation are given by
\begin{eqnarray}\label{eq:Tpopr}
\Delta T &\geq& \frac{1}{\alpha} \sqrt{\frac{(1-\eta)(1+2N)}{4\eta\bar{N}}} + \frac{(1 - \eta) \hbar \omega \bar{N}}{M C_s}, \\
\label{eq:Tcohpopr}
\Delta T &\geq& \frac{1}{\alpha} \sqrt{\frac{1+2(1-\eta)N}{4\eta\bar{N}}} + \frac{(1 - \eta) \hbar \omega \bar{N}}{M C_s}
\end{eqnarray}
for the squeezed-vacuum and coherent states, respectively. From the above formulas, we see that there is a maximum value $\bar{N}_{\rm max}$ of $\bar{N}$ for which the error is minimal, and increasing $\bar{N}$ from that point on would only decrease the precision because of the heating of the sample. This behavior is clearly visible in Fig.~\ref{fig:bounds} (which was prepared for the PPKTP crystal; all necessary parameters for this material are given in the figure's caption), where for moderately small $\bar{N}$ the error decreases, but eventually it starts to increase when growing $\bar{N}$ disturbs the sample's temperature. Fortunately, this maximum value $\bar{N}_{\rm max}$ is of the order of $10^{13}$--$10^{14}$ photons and, thus, the interferometric thermometers are noninvasive for most of the realistic probe states.

\subsection{Comparison with noninvasive classical thermometers}\noindent
We compare the precision offered by our interferometric thermometer with the theoretical limit imposed on the idealized pyrometer that we found in Sec.~\ref{sec:pyrometer}. We perform this comparison for the PPKTP crystal, which is commonly used in many quantum optics experiments, and its basic physical properties can be easily found \cite{Nikogosyan2005}. The results of our calculations for the PPKTP crystal near $T=298$ K are presented in Fig.~\ref{fig:bounds}. As can be easily checked, the best precision achieved by our interferometric approach is $\Delta T \approx 1.4$ nK for light with wavelength $\lambda=1064$ nm and $\bar{N} = 3.2 \times 10^{13}$ photons prepared in the squeezed-vacuum state. This is much higher than the precision of the commercially available pyrometers and even higher than the theoretical limit for the idealized pyrometer of $\Delta T = 452$ nK. Naturally, a squeezed-vacuum state containing that many photons would be very hard to produce \footnote{In fact, the best known squeezed-vacuum beams contain approximately 10 photons \cite{LIGO2013}, in which case the precision of our thermometer, being limited to around 0.1 K (see Fig.~\ref{fig:bounds}), would be comparable to that of the commercially available pyrometers.}. However, even for coherent input states, the precision of the interferometric thermometer is still excellent, with $\Delta T \approx 24$ nK obtained with $\bar{N} = 5.6 \times 10^{14}$ photons (this number of photons at $\lambda=1064$ nm would correspond to a laser pulse energy of 0.1 mJ and the average power of 100 mW at the repetition rate of 1 kHz, which is easily accessible with current technology). Therefore, we find that the interferometric thermometry does not need to use specially designed quantum states of light to surpass the classical optical thermometers.

One could argue that our analysis is not comprehensive because we have not fully exploited all possible sources of information about the temperature in our setup. For example, we could have additionally tried to measure the mean number of thermal photons $N$ being radiated by the sample at frequency $\omega$, which clearly is temperature dependent. Furthermore, it very much depends on the type of the sample, but very often the transmissivity $\eta$ also depends on the temperature. Therefore, we could have tried to estimate the temperature from $\eta$ as well. The QFIs for estimation of $N$ and $\eta$ were already calculated in Ref.~\cite{Monras2011} and, as we found, the estimation of these additional parameters provides negligible information about temperature when compared with the information obtained from the estimation of the optical phase. Moreover, if, in spite of that, we still wished to estimate the mean number of thermal photons $N$, then it would be optimal to send a vacuum state through the sample \cite{Monras2011}, which is equivalent to measuring the thermal radiation emitted by the sample at a \textit{single} frequency $\omega$, and interestingly provides information that we have already found in Eq.~(\ref{eq:singleFisher}) in the section devoted to the idealized pyrometers.

\section{Discussion and future directions}\label{sec:conclusions}\noindent
In summary, we provided a detailed description of the quantum interferometric thermometer by analyzing the interaction between the single-mode Gaussian states of light and a heated sample modeled as a dissipative thermal reservoir. We found that the single-mode squeezed-vacuum state is optimal for temperature measurement, offering precision in the nanokelvin range; however, coherent input states provide an excellent performance as well. Moreover, we also found a very elegant formula given by Eq.~(\ref{eq:pyrometer}) that lower bounds the fundamental precision of the idealized pyrometer, which infers the temperature from a measurement of the total thermal radiation emitted by a heated sample. The interferometric thermometer provides a superior performance in temperature sensing even when compared with this idealized pyrometer. One of the main advantages of interferometric thermometry is its noninvasiveness, as highly precise temperature measurements can be obtained without disturbing the sample's temperature. We believe this feature to be crucial and hope that interferometric thermometers could be used, for example, for active temperature stabilization of nonlinear crystals such as the PPKTP crystal. We note, however, that further studies are required to access the true performance of this method for temperature measurements in nonlinear crystals. Those studies would need to take into account intrinsic noise sources such as the Brownian and the thermodynamical noise sources that may be important for specific materials and setup geometries \cite{Braginsky1999}.

Apart from nonlinear crystals, we predict that interferometric thermometers should prove very useful for temperature measurements of atomic vapors. At the moment, the most popular optical technique for temperature measurements in gaseous mediums such as atomic vapors is the absorption spectroscopy in which a laser light is shone on an atomic-vapor cell producing the absorption spectrum. In order to estimate the temperature of atomic vapor, it is then necessary to fit the observed spectrum to a theoretical model described by the so-called Voigt profile, which accurately models the intensity profile of absorption spectral lines by including the contributions of natural linewidth and Doppler broadening \cite{Chrapkiewicz2012, Demtroder2014}. This fitting procedure normally involves a prior knowledge of atomic-vapor parameters, such as the density of the vapor, which has a complicated dependence on the temperature. Because of this, in practice, the precision of this method is limited to, at most, $\pm$0.1 K \cite{Truong2011, Stace2012}, which is rather poor \footnote{Although, an important experimental advancement in this area was recently achieved \cite{Truong2015}.}. Furthermore, if we try to estimate the temperature of an atomic vapor by scanning the whole absorption spectrum, then we will very likely disturb its state when we probe it around resonant transitions \cite{Radek2014}. This is very problematic if we wish to use our atomic-vapor cell as a quantum memory. Therefore, it is clear that an alternative, simpler method of temperature measurements for atomic vapors could be useful. We believe that our noninvasive interferometric approach is ideally suited for this kind of application and it would be of great interest to determine the fundamental precision offered by the interferometric thermometer for temperature measurements in realistic atomic vapors.

Finally, we note that interferometric thermometers operating at the telecom wavelengths of 1550 nm can also be readily used for temperature measurements of silicon semiconductor electronic devices, as silicon is almost transparent at these longer wavelengths.

\section{Acknowledgements}
We would like to thank Rafa{\l} Demkowicz-Dobrza{\'n}ski and Rados{\l}aw Chrapkiewicz for many helpful discussions and comments. This work was supported by the European Union Seventh Framework Programme (FP7/2007-2013) projects SIQS (Grant agreement No. 600645; co-financed by the Polish Ministry of Science and Higher Education) and PhoQuS@UW (Grant agreement No. 316244).

\appendix
\section*{Appendix}
\setcounter{section}{1}\noindent
In order to calculate the SLD and the QFI for the optical phase, we adopt a technique presented in Ref.~\cite{Monras2010}; for the sake of convenience, we adopt most of the notation used in that work. We recall that according to the master equation give in Eq.~(\ref{eq:master}), after a time $t$, an initial state $\rho_{0}$ evolves to $\rho_\varphi = \exp[\mathcal{G}(\omega, N, \Gamma) t]\rho_{0} = \mathcal{S}(\varphi, N, \gamma) \rho_{0}$, where $\mathcal{S}(\varphi, N, \gamma) = \exp[\mathcal{G}(\varphi, N, \gamma)]$, and here $\gamma = \Gamma t$ is the effective coupling to a thermal reservoir (which is related to the photon loss coefficient $\eta$ via the relation $\eta = e^{-\gamma}$) and $\varphi = \omega t$ is the optical phase we wish to estimate. In the following derivation, we will drop the dependence of the superoperators $\mathcal{G}$ and $\mathcal{S}$ on the mean number of photons $N$ and the photon loss $\gamma$ because we are only concerned with finding the SLD for the phase.

We begin by finding the partial derivative of $\rho_\varphi$ with respect to $\varphi$, which can be neatly expressed as
\begin{equation}\label{eq:chain}
\partial_{\varphi} \rho_\varphi = \partial_{\varphi} \mathcal{S}(\varphi) \rho_{0} = \partial_{\varphi} \exp[\mathcal{G}(\varphi)] \rho_{0}.
\end{equation}
Using the following relation \cite{Monras2010}:
\begin{eqnarray}
\partial_{\varphi} \exp[\mathcal{G}(\varphi)] &=& \int_{0}^{1} e^{u \mathcal{G}(\varphi)} \partial_{\varphi} \mathcal{G}(\varphi) e^{(1-u)\mathcal{G}(\varphi)} du \nonumber \\
&=&  \int_{0}^{1} e^{u \mathcal{G}(\varphi)} \partial_{\varphi} \mathcal{G}(\varphi) e^{-u \mathcal{G}(\varphi)} du \, \mathcal{S}(\varphi) \nonumber \\
\end{eqnarray}
we obtain
\begin{equation}
\partial_{\varphi} \rho_\varphi = \mathcal{D}_{\varphi} \mathcal{S}(\varphi) \rho_{0} = \mathcal{D}_{\varphi} \rho_\varphi,
\end{equation}
where the superoperator $\mathcal{D}_{\varphi}$ is given by
\begin{equation}
\mathcal{D}_\varphi = \int_{0}^{1} e^{u \mathcal{G}(\varphi)} \partial_{\varphi} \mathcal{G}(\varphi) e^{-u \mathcal{G}(\varphi)} du,
\end{equation}
which can be easily calculated using the Baker-Campbell-Hausdorff formula, resulting in $\mathcal{D}_{\varphi} = -iH$.

Recalling the definition of the SLD, $\partial_{\varphi}\rho_\varphi = L_{\varphi} \circ \rho_\varphi$, we can write
\begin{equation}\label{eq:Sylvester}
\beta_{ij} [\chi^i\chi^j,\,\rho_\varphi] = L_{\varphi} \circ \rho_\varphi,
\end{equation}
where we have introduced $\boldsymbol{\chi} = (a,\,a^\dagger)$ and
\begin{equation}
\beta = \left(\begin{array}{cc}
0 & - i/2\\
- i/2 & 0
\end{array}\right)
\end{equation}
and the Einstein summation convention is assumed. Equation~(\ref{eq:Sylvester}) is an example of the Sylvester equation $Y = Z \circ X$ \cite{Bhatia1997}, which has the following formal solution:
\begin{equation}\label{eq:solutionofSylvester}
L_{\varphi} = 2\int_0^\infty e^{-v \rho_\varphi}\beta_{ij}[\chi^i\chi^j,\,\rho_\varphi]e^{-v \rho_\varphi} dv.
\end{equation}
In the next step, we introduce $\tilde{\chi}^i = \chi^i - \langle\chi^i\rangle$ and calculate
\begin{widetext}
\begin{eqnarray}\label{eq:BlaBla}
\int_{0}^{\infty}e^{-v\rho_\varphi}\tilde{\chi}^{i}\tilde{\chi}^{j}\rho_\varphi e^{-v\rho_\varphi}dv &=&
\int_{0}^{\infty}e^{-v\rho_\varphi}\tilde{\chi}^{i}e^{v\rho_\varphi}e^{-v\rho_\varphi}\tilde{\chi}^{j}e^{v\rho_\varphi}e^{-v\rho_\varphi}\rho_\varphi e^{-v\rho_\varphi}dv \nonumber \\
&=& \sum_{m,n}\frac{(-1)^{m+n}}{m!n!} [(\mathbb{F}-\mathbb{1})^{m}]_{k}^{i}[(\mathbb{F}-\mathbb{1})^{n}]_{l}^{j}\tilde{\chi}^{k}\rho_\varphi^{m}\tilde{\chi}^{l}\rho_\varphi^{n+1} \int_{0}^{\infty} v^{m+n} e^{-2v\rho_\varphi} dv \nonumber \\
&=&\sum_{m,n}\frac{(-1)^{m+n}}{2^{m+n+1}}{m+n \choose n}[(\mathbb{F}-\mathbb{1})^{m}]_{k}^{i}[(\mathbb{F}-\mathbb{1})^{n}]_{l}^{j}\tilde{\chi}^{k}\rho_\varphi^{m}\tilde{\chi}^{l}\rho_\varphi^{-m} \nonumber \\
&=&\sum_{m,n}\frac{(-1)^{m+n}}{2^{m+n+1}}{m+n \choose n}[(\mathbb{F}-\mathbb{1})^{m}]_{k}^{i}[(\mathbb{F}-\mathbb{1})^{n}]_{l}^{j}\tilde{\chi}^{k}[\mathbb{F}^{m}]_{s}^{l}\tilde{\chi}^{s} \nonumber \\
&=&\sum_{m,n}\frac{(-1)^{m+n}}{2^{m+n+1}}{m+n \choose n}\{[(\mathbb{F}-\mathbb{1})\otimes\mathbb{F}]^{m}[\mathbb{1}\otimes(\mathbb{F}-\mathbb{1})]^{n}\}_{kl}^{ij}\tilde{\chi}^{k}\tilde{\chi}^{l} \nonumber \\
&=&\sum_{q=0}^{\infty}\frac{(-1)^{q}}{2^{q+1}}\sum_{n=0}^{q}{q \choose n}\{[(\mathbb{F}-\mathbb{1})\otimes\mathbb{F}]^{q-n}[\mathbb{1}\otimes(\mathbb{F}-\mathbb{1})]^{n}\}_{kl}^{ij}\tilde{\chi}^{k}\tilde{\chi}^{l} \nonumber \\
&=&\sum_{q=0}^{\infty}\frac{(-1)^{q}}{2^{q+1}}\{[(\mathbb{F}-\mathbb{1})\otimes\mathbb{F}+\mathbb{1}\otimes(\mathbb{F}-\mathbb{1})]^{q}\}_{kl}^{ij}\tilde{\chi}^{k}\tilde{\chi}^{l} \nonumber \\
&=&[(\mathbb{F}\otimes\mathbb{F}+\mathbb{1}\otimes\mathbb{1})^{-1}]_{kl}^{ij}\tilde{\chi}^{k}\tilde{\chi}^{l},
\end{eqnarray}
\end{widetext}
where $\mathbb{F} = H^{\dagger}f(\tilde{\Sigma})H$, with $H$ being the Hadamard matrix,
\begin{equation}
H = \frac{1}{\sqrt{2}}\left(\begin{array}{cc}
1 & 1\\
-i & i
\end{array}\right),
\end{equation}
and $f(x)=\frac{x-i/2}{x+i/2}$ and $\tilde{\Sigma} = \Sigma\Omega$, where $\Sigma$ is the covariance matrix of the output state $\rho_{\varphi}$ with $\Omega$ begin the symplectic matrix,
\begin{equation}
\Omega=\left(\begin{array}{cc}
0 & 1\\
-1 & 0
\end{array}\right).
\end{equation}
In the derivation of Eq.~(\ref{eq:BlaBla}), we used a number of identities, which can all be found in Appendix C of Ref.~\cite{Monras2010}: (i) in the second line, we used
\begin{equation}
e^{-v\rho_\varphi}\tilde{\chi}^{i}e^{v\rho_\varphi} = \sum_{m}\frac{(-1)^{m}}{m!}v^{m} [(\mathbb{F}-\mathbb{1})^{m}]_{k}^{i}\tilde{\chi}^{k}\rho_\varphi^{m};
\end{equation}
(ii) in the fourth line, we used $\rho_\varphi^{m}\tilde{\chi}^{l}\rho_\varphi^{-m}=[\mathbb{F}^{m}]_{s}^{l}\tilde{\chi}^{s}$; (iii) and in the sixth line, we changed the variables $m + n = q$ and replaced $\sum_{m,n = 0}^{\infty}$ with $\sum_{q = 0}^{\infty}$ $\sum_{n = 0}^{q}$.  Similarly, we can show that
\begin{eqnarray}\label{eq:BlaBla2}
&&\int_{0}^{\infty}e^{-v\rho_\varphi}\rho_\varphi \tilde{\chi}^{i}\tilde{\chi}^{j}e^{-v\rho_\varphi}dv \nonumber \\
&&= [(\mathbb{F}\otimes\mathbb{F}+\mathbb{1}\otimes \mathbb{1})^{-1}(\mathbb{F}\otimes\mathbb{F})]_{kl}^{ij}\tilde{\chi}^{k}\tilde{\chi}^{l}.
\end{eqnarray}
Replacing now $\chi^i$ with $\chi^i = \tilde{\chi}^i - \langle\chi^i\rangle$ in Eq.~(\ref{eq:solutionofSylvester}) and using Eqs.~(\ref{eq:BlaBla}) and (\ref{eq:BlaBla2}), we obtain an expression for the SLD for the optical phase $\varphi$,
\begin{eqnarray}
L_\varphi&=&2\beta_{ij}[\mathbb{F}\otimes\mathbb{F}+\mathbb{1}\otimes \mathbb{1})^{-1}(\mathbb{1}\otimes \mathbb{1}-\mathbb{F}\otimes\mathbb{F})]^{ij}_{kl}\tilde{\chi}^k\tilde{\chi}^l \nonumber \\
&+&[(\mathbb{1}+\mathbb{F})^{-1}-(\mathbb{1}+\mathbb{F}^{-1})^{-1}]^i_k\tilde{\chi}^k\langle\chi^j\rangle \nonumber \\
&+&[(\mathbb{1}+\mathbb{F})^{-1}-(\mathbb{1}+\mathbb{F}^{-1})^{-1}]^j_k\tilde{\chi}^k\langle\chi^i\rangle,
\end{eqnarray}
which, after a very tedious but rather straightforward algebra, can be further simplified to
\begin{eqnarray}
\nonumber L_\varphi &=& i \tilde{\beta}_{ij} \Big([D^{-1}(\tilde{\Sigma}\otimes \mathbb{1}+\mathbb{1}\otimes\tilde{\Sigma})]^{ij}_{kl}(\tilde{R}^k \circ \tilde{R}^l + i\Omega^{kl}/2) \nonumber \\
&+&(\tilde{\Sigma}^{-1})^i_k\langle R^j\rangle\tilde{R}^k+(\tilde{\Sigma}^{-1})^j_k\langle R^i\rangle\tilde{R}^k\Big),
\end{eqnarray}
where we have defined
\begin{equation}
D = \tilde{\Sigma}\otimes\tilde{\Sigma}-\frac{1}{4}\mathbb{1}\otimes \mathbb{1}
\end{equation}
and $\tilde{\beta}_{ij}=\beta_{qp}(H^{\dagger}\otimes H^{\dagger})^{qp}_{ij}$, which is explicitly given by
\begin{equation}
\tilde{\beta} = \left(\begin{array}{cc}
-i/2 & 0\\
0 & -i/2
\end{array}\right)
\end{equation}
and $R^k=H^k_l\chi^l$ and, similarly, $\tilde{R}^k=H^k_l\tilde{\chi}^l$. The last transformation means that ${\bf R} = (x, p)$ [$\tilde{\bf{R}} = (\tilde{x}, \tilde{p})$], where $x$ and $p$ are the canonical position and momentum operators [$\tilde{x}$ and $\tilde{p}$ are the displaced canonical position and momentum operators].

In the final step, we rewrite the SLD for the optical phase as
\begin{equation}\label{eq:SLDphi}
L_{\varphi} = \frac{4(\Sigma_{22} - \Sigma_{11})}{1 + 4\Sigma_{11}\Sigma_{22}} \tilde{x} \circ \tilde{p} + \frac{\bar{p}}{\Sigma_{11}} \tilde{x} - \frac{\bar{x}}{\Sigma_{22}} \tilde{p},
\end{equation}
where $\Sigma_{11}$ and $\Sigma_{22}$ are the diagonal elements of the covariance matrix $\Sigma$ of the output state $\rho_{\varphi}$ given in Eq.~(\ref{eq:Sigma}), and $\bar{x}$ and $\bar{p}$ are the mean displacements in the respective canonical position and momentum quadratures given in Eqs.~(\ref{eq:meanx}) and (\ref{eq:meanp}), all of which are calculated with $\varphi = 0$. At this point, we need to make two comments: (i) the SLD for the optical phase has zero expectation, that is, ${\rm Tr}[\rho_\varphi L_{\varphi}] = 0$, as any legitimate SLD should, and (ii) it is expressed in terms of the first and second moments of the output Gaussian state $\rho_\varphi$ with $\varphi = 0$ because, in our setup, the QFI for the optical phase $\varphi$ is independent of the actual value of $\varphi$; hence, we choose to present the SLD in its simplest form.

Given the above SLD, we can calculate the corresponding QFI for the optical phase via
\begin{equation}
Q_{\varphi} = {\rm Tr}[\rho_\varphi L^{2}_{\varphi}],
\end{equation}
which is given in Eq.~(\ref{eq:QFIphase}) in the main text.

\end{document}